\documentclass[10pt,A4paper,twocolumn]{article}
\usepackage{amsmath}
\usepackage{graphicx}
\usepackage{fancyhdr}
\usepackage{overcite}
\usepackage[left=2cm,top=1cm,right=2cm]{geometry}

\begin{document}
\twocolumn[
\begin{center}
\begin{Large}
\bf{A spatial refractive index sensor using whispering gallery modes
in an optically trapped microsphere}
\end{Large}
\vskip 0.5cm {Peter Zijlstra$^{a)}$, Karen L. van der Molen,
Allard P. Mosk$^{b)}$} \vskip 0.5 cm \textit{Complex Photonic
Systems, MESA$^+$
Institute for Nanotechnology and Department of Science and Technology\\
University of Twente, PO Box 217, 7500 AE Enschede, The
Netherlands.} \\
\vskip 0.2 cm \small{ a) Current address: Centre for
Mirco-Photonics, Swinburne University of Technology, Melbourne,
Australia.\\
b) E-mail: a.p.mosk@utwente.nl} \vskip 0.5 cm
\end{center} 

\textbf{\small{We propose the use of an optically trapped, dye
doped polystyrene microsphere for spatial probing of the
refractive index at any position in a fluid. We demonstrate the
use of the dye embedded in the microsphere as an internal
broadband excitation source, thus eliminating the need for a
tunable excitation source. We measured the full width at half
maximum of the TE and TM resonances, and their frequency spacing
as a function of the refractive index of the immersion fluid. From
these relations we obtained an absolute sensitivity of $5 \times
10^{-4}$ in local refractive index, even when the exact size of
the microsphere was not \emph{a priori}\ known. }}
\vskip 0.5 cm ]

Microspheres can act as high Q resonators in the optical regime
\cite{vahala03,kalkman06} and can store electromagnetic waves by
repeated total internal reflection. Constructive interference
results in an enhanced internal field which is called a whispering
gallery mode (WGM). Because of the evanescent interaction between
the WGM and the surrounding medium the WGM's resonance frequency
is sensitive to changes in the refractive index of the sphere's
surroundings. This sensitivity makes microspheres an useful tool
in sensory applications. Recently Hanumegowda \textit{et al.} demonstrated
the use of microspheres in refractometric sensing
\cite{hanumegowda05}. Moreover Vollmer \textit{et al.} \cite{vollmer02} and
White \textit{et al.} \cite{white05} demonstrated protein detection using
microspheres. In all these studies
\cite{hanumegowda05,vollmer02,white05} narrowband light from a
tunable diode laser was evanescently coupled to the microsphere
via an optical fiber. Spatial mapping of the refractive index was
not possible because the microsphere could not be freely moved
through the sample.

In this letter we demonstrate the use of a dye-doped polystyrene
microsphere as a refractometric sensor. The polystyrene
microsphere was trapped with optical tweezers, by which it could
be positioned anywhere within the fluid. It can therefore be used
to sense the refractive index in spatially inhomogeneous media,
for example near interfaces or in mixing flows. In this system we
did not need a tunable laser to excite WGMs, instead we used the
dye embedded in the microsphere itself as a broadband excitation
source. The absolute frequency of the WGM, which was used to probe
the refractive index in previous studies
\cite{hanumegowda05,vollmer02,white05}, is very sensitive to the
size of the sphere. Microspheres synthesized in bulk always have a
considerable size polydispersity, and therefore a wide
distribution of absolute resonance frequencies. Measurements of
the absolute resonance frequency of a particle that has not been
individually calibrated does not provide information about the
refractive information of the medium. In contrast, we show that
both the full width at half maximum (FWHM) of the emission peaks
and the frequency spacing between adjacent transverse electric (TE) and transverse magnetic (TM) modes are
sensitive and robust probes of the refractive index of the
sphere's surroundings. We show that a calibration of the
individual microspheres is not needed in this case.

The FWHM of a WGM is determined by the loss upon total internal
reflection at the curved surface of the microsphere. These losses
depend on the mode number $l$ and the refractive index contrast
$m=m_0/m_I$, where $m_0$ and $m_I$ are the refractive indices of
the immersion medium and the microsphere respectively. The FWHM of
a WGM can be expressed as \cite{lam92}
\begin{equation}
\frac{\Gamma}{2}=[Nx^2n_l(x)^2]^{-1}, \label{FWHM_lam}
\end{equation}
with $n_l$ the spherical Neumann function, $x=m_0ka$ the size
parameter at which the WGM occurs, where $k$ is the wave vector in
vacuum and $a$ the radius of the microsphere, and
\begin{equation}
N=
\begin{cases}
m^2-1 & \text{for TE modes} \\
(m^2-1)[\mu^2+(\mu^2/m^2-1)] & \text{for TM modes,}
\end{cases}
\end{equation}
where $\mu=\nu/x$, with $\nu=l+1/2$.

The frequency spacing between a TE and TM mode of the same mode
number is due to a difference in phase shift of s and p polarized
light upon reflection at the sphere surface. The absolute frequency
at which a WGM occurs is defined as the real part of the pole of the
scattering coefficient \cite{bh83}, and we calculated the frequency
spacing by numerically calculating the poles for TE and TM modes.

In our experiments we used a suspension of dye-doped polystyrene
microspheres in water (G1000, Duke Scientific Corporation). The
manufacturer specified a mean radius of 5.0 $\pm$ 0.3 $\mu$m
(standard deviation). We measured the size distribution of 24
microspheres and found a mean radius of 5.3 $\mu$m with a maximum
deviation of only 0.5 \%. Taking into account a conservative error
margin, we assumed that the size of any sphere we trap was 5.30
$\pm$ 0.05 $\mu$m. The microspheres were doped with 2\% dye which
emits in the green part of the spectrum ($\lambda=480-540$ nm). The
sample holder was a borosilicate glass capillary with inner
dimensions $0.1\times2\times40$ mm$^3$.

With optical tweezers operating at 1064 nm we held a single
microsphere far away from any surface. The excitation light source
was a continuous wave Argon laser (Spectra Physics Satellite 2016)
emitting at 488 nm with a maximum output power of 30 mW, attenuated
to 500 $\mu$W on the sample. The laser light was slightly defocused
on the sphere, so that the whole sphere was illuminated. The
emission of the dye was collected through a water immersed objective
(NA 1.2) and detected by an ICCD (Princeton Instruments ICCD 576),
connected to a spectrograph (Oriel Instruments, MS257) which has a
Gaussian response function with a width of 4.5 cm$^{-1}$.

Tuning of the refractive index of the immersion fluid $m_0$ was
achieved by mixing water with ethylene glycol ($m_0$ = 1.4317 at 293
K and 589 nm) in various fractions. Ethylene glycol does not
chemically affect polystyrene and is miscible with water in all
volume fractions. Refractive indices of the mixtures were measured
with an Abbe refractometer type 1T which is temperature stabilized
and has an accuracy of 2 $\times 10^{-4}$ in refractive index.

\begin{figure}[]
    \begin{center}
      \includegraphics{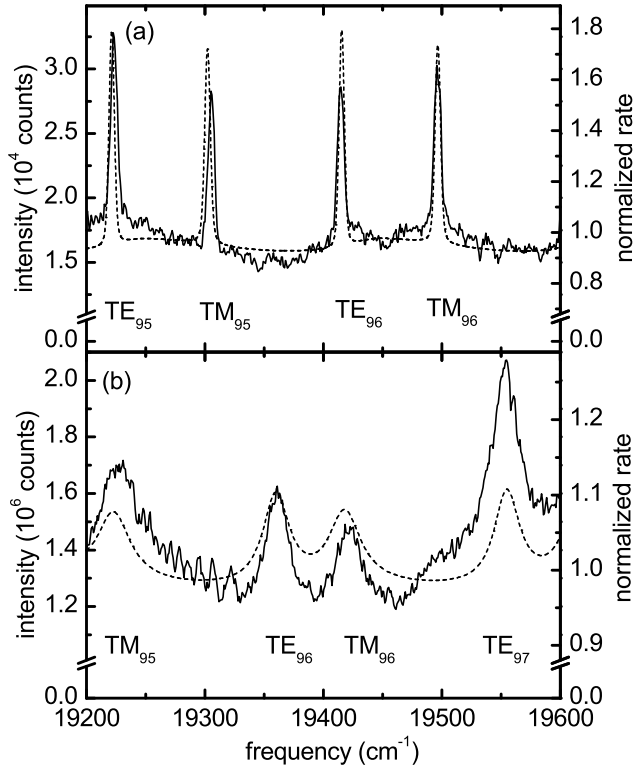}
    \end{center}
   \caption{\label{sp_cw_133139} \small{Measured emission intensity (solid lines) of single dye
doped polystyrene microspheres immersed in (a) water ($m_0$ =
1.3325) and (b) a mixture of water and ethylene glycol ($m_0$ =
1.3978). The dashed lines show the normalized radiation rates as
predicted by EMT for (a) $a$=5.33 $\mu$m, and (b) $a$=5.319 $\mu$m.}}
\end{figure}

We compared our experimental emission spectra to an elaboration of
Mie theory (EMT) developed by Chew \textit{et al.} describing the emission
of dipoles inside a spherical cavity \cite{chew87,chew88}. The only
parameters were the refractive index of the sphere $m_I$, the
refractive index of the immersion medium $m_0$, and the sphere
radius $a$. The refractive index of the sphere was measured to be
$m_I$ = 1.586 - i $\cdot$ 1.5 $\times$ 10$^{-5}$ \cite{molen06}. The
refractive index of the immersion fluid $m_0$ was measured with the
refractometer. The only free parameter in our system was the radius
of the sphere $a$. The theoretical emission spectra were corrected
for the response function of the spectrograph.

In Fig. \ref{sp_cw_133139} we show the emission spectra obtained
from a single dye doped polystyrene microsphere immersed in (a)
water ($m_0$ = 1.3325) and (b) a mixture of water and ethylene
glycol ($m_0$ = 1.3978). Sharp peaks were observed in the spectrum
due to WGMs. The dashed lines show the radiation rates (normalized
to the rate in bulk polystyrene) as predicted by EMT for (a) $a$ =
5.33 $\mu$m, and (b) $a$ = 5.319 $\mu$m. Comparing the spectra in
(a) and (b) we observed that the decrease in refractive index
contrast induced an increase in FWHM of the WGMs and a decrease in
the spacing between the adjacent TE and TM modes.

As we have shown in Fig. \ref{sp_cw_133139} it is possible to obtain
$m_0$ and $a$ by fitting the experimental spectra with a theoretical
emission spectrum. However, the initial values for the parameters
must be chosen rather carefully. The same parameters can be obtained
in a more straightforward way by comparing only the FWHM and the
TE$_{96}$-TM$_{96}$ mode spacing to predictions from EMT. Both the
FWHM and the mode spacing were determined by fitting a Voigt profile
\cite{loudon00} to an individual WGM in the measured emission
spectrum. The Voigt profile models the convolution of the Lorentzian
width of the resonance and the Gaussian response function of the
spectrograph. To accurately fit the resonance shape, we subtracted a
quadratic fit of the background intensity.

\begin{figure}[]
    \begin{center}
      \includegraphics{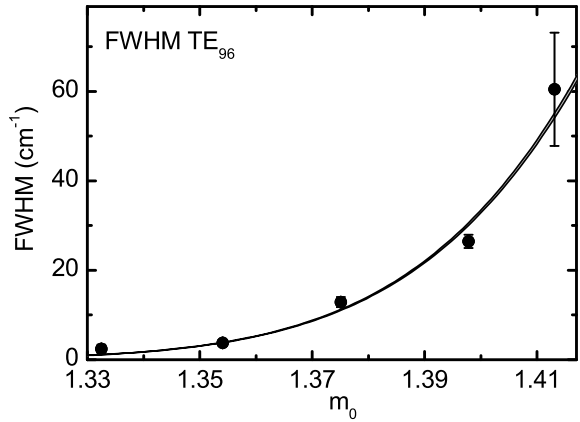}
    \end{center}
   \caption{\label{fwhm_TE96} \small{Measured Lorentzian FWHM versus $m_0$ for the TE$_{96}$
mode. The solid lines show the theoretical FWHM for the biggest and
the smallest spheres in the measured size distribution, and was
calculated from Eq. (\ref{FWHM_lam}).}}
\end{figure}

In Fig. \ref{fwhm_TE96} we show the Lorentzian FWHM as a function of
$m_0$ for the TE$_{96}$ mode. The size of the error bars was taken
from the error in the Lorentzian width of the fits. The large error
at $m_0$ = 1.41 was due to a low signal to noise ratio resulting
from a low refractive index contrast. The theoretical FWHMs (solid
lines) were calculated for the biggest and smallest spheres in the
measured size distribution, and were calculated from Eq.
(\ref{FWHM_lam}). It can be seen that the FWHM is very insensitive
to the sphere size as the calculated curves are almost identical.

\begin{figure}[]
    \begin{center}
      \includegraphics{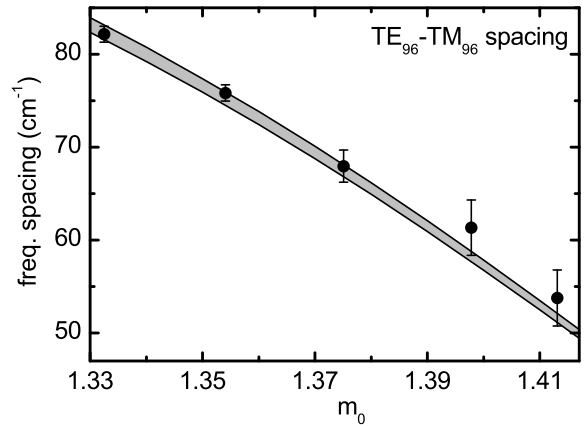}
    \end{center}
   \caption{\label{spacing_fig} \small{Measured TE$_{96}$-TM$_{96}$ mode spacing versus $m_0$.
The grey area shows the theoretical mode spacing calculated for the
biggest and smallest spheres in the measured size distribution.}}
\end{figure}

In Fig. \ref{spacing_fig} we show the experimental
TE$_{96}$-TM$_{96}$ mode spacing, obtained from the central
frequency of the fitted Voigt line shapes. The error bars are the
uncertainty in the central frequency of the Lorentzian fits. The
theoretical mode spacing is again calculated for the biggest and
smallest spheres in the measured size distribution, and the result
is shown as a grey area. It can be seen that the TE$_{96}$-TM$_{96}$
mode spacing is slightly more sensitive to sphere size than the FWHM
(Fig.\ref{fwhm_TE96}).

Figs. \ref{fwhm_TE96} and \ref{spacing_fig} show that our
measurements correspond very well to the theoretical model, even
though the size of the trapped microsphere varies between
measurements due to size polydispersity. The polydispersity in sphere size introduced an inaccuracy in the theoretical values for
the FWHM and TE-TM mode spacing of $<$ 1 \%.

One can use either the FWHM or the mode spacing to measure the
refractive index of the medium. However, the sensitivity and
robustness of these methods is not the same. The
TE$_{96}$-TM$_{96}$ mode spacing is nearly linear in our
refractive index range with a slope of 385 cm$^{-1}$ per
refractive index unit. An accuracy in width of 2 cm$^{-1}$ was
easily achieved for refractive indices below 1.39, resulting in a
sensitivity of nearly 5 $\times$ 10$^{-3}$ in refractive index.
When measuring the FWHM of a WGM the sensitivity of the sensor is
higher, and varies throughout our refractive index domain. From
the slope of the curve in Fig. \ref{fwhm_TE96} we obtained a
sensitivity of 5 $\times$ 10$^{-4}$ in refractive index around
$m_0$ = 1.41. The FWHM method leads to the most precise
measurements in the ideal case, however, this method is very
sensitive to contamination at the sphere surface. The mode spacing
is a more robust method and may therefore be used when
contamination cannot be excluded.

In conclusion we demonstrated a spatial probe for the refractive
index comprising an optically trapped dye doped microsphere. The
use of optical tweezers gives us the freedom to measure the
refractive index anywhere in the fluid. This freedom allows for
spatial mapping of refractive index gradients or inhomogeneities
caused by, for example, flow in microfluidic devices or chemical
processes. The emission of an embedded dye acts as the internal
source to excite the WGMs in the microsphere. We measured the FWHM
and the frequency spacing between TE and TM modes. The measured
values show excellent agreement with theoretical predictions,
confirming the suitability of the system for spatial probing of
the refractive index without the need for individual calibration.
Very recently, Kn\"{o}ner \textit{et al.} \cite{knoner} have demonstrated a
method to determine the refractive index of a trapped microsphere
by measuring the trapping force. Their and our methods are
complimentary and could easily be combined, providing a versatile
system for sensory applications.

\section*{Acknowledgements}
We thank Ad Lagendijk for useful discussions and support and
L\'{e}on Woldering for his help with the sample preparation. This work
is part of the research program of the 'Stichting voor Fundamenteel
Onderzoek der Materie' (FOM), which is financially supported by the
'Nederlandse Organisatie voor Wetenschappelijk Onderzoek' (NWO).

\end{document}